\documentclass[useAMS,usenatbib]{mnras}
\usepackage{multirow}
\usepackage{epsfig}
\usepackage{subfigure}
\usepackage{graphicx}	
\usepackage{amsmath}	
\usepackage{amssymb}	
\usepackage{multicol}        
\usepackage{bm}		
\usepackage{pdflscape}	

\makeatletter
\let\jnl@style=\rm
\def\ref@jnl#1{{\jnl@style#1}}

\def\aj{\ref@jnl{AJ}}                   
\def\araa{\ref@jnl{ARA\&A}}             
\def\apj{\ref@jnl{ApJ}}                 
\def\apjl{\ref@jnl{ApJ}}                
\def\apjs{\ref@jnl{ApJS}}               
\def\ao{\ref@jnl{Appl.~Opt.}}           
\def\apss{\ref@jnl{Ap\&SS}}             
\def\aap{\ref@jnl{A\&A}}                
\def\aapr{\ref@jnl{A\&A~Rev.}}          
\def\aaps{\ref@jnl{A\&AS}}              
\def\azh{\ref@jnl{AZh}}                 
\def\baas{\ref@jnl{BAAS}}               
\def\jrasc{\ref@jnl{JRASC}}             
\def\memras{\ref@jnl{MmRAS}}            
\def\mnras{\ref@jnl{MNRAS}}             
\def\pra{\ref@jnl{Phys.~Rev.~A}}        
\def\prb{\ref@jnl{Phys.~Rev.~B}}        
\def\prc{\ref@jnl{Phys.~Rev.~C}}        
\def\prd{\ref@jnl{Phys.~Rev.~D}}        
\def\pre{\ref@jnl{Phys.~Rev.~E}}        
\def\prl{\ref@jnl{Phys.~Rev.~Lett.}}    
\def\pasp{\ref@jnl{PASP}}               
\def\pasj{\ref@jnl{PASJ}}               
\def\qjras{\ref@jnl{QJRAS}}             
\def\skytel{\ref@jnl{S\&T}}             
\def\solphys{\ref@jnl{Sol.~Phys.}}      
\def\sovast{\ref@jnl{Soviet~Ast.}}      
\def\ssr{\ref@jnl{Space~Sci.~Rev.}}     
\def\zap{\ref@jnl{ZAp}}                 
\def\nat{\ref@jnl{Nature}}              
\def\iaucirc{\ref@jnl{IAU~Circ.}}       
\def\aplett{\ref@jnl{Astrophys.~Lett.}} 
\def\apspr{\ref@jnl{Astrophys.~Space~Phys.~Res.}}
\def\bain{\ref@jnl{Bull.~Astron.~Inst.~Netherlands}}
\def\fcp{\ref@jnl{Fund.~Cosmic~Phys.}}  
\def\gca{\ref@jnl{Geochim.~Cosmochim.~Acta}}   
\def\grl{\ref@jnl{Geophys.~Res.~Lett.}} 
\def\jcp{\ref@jnl{J.~Chem.~Phys.}}      
\def\jgr{\ref@jnl{J.~Geophys.~Res.}}    
\def\jqsrt{\ref@jnl{J.~Quant.~Spec.~Radiat.~Transf.}}
\def\memsai{\ref@jnl{Mem.~Soc.~Astron.~Italiana}}
\def\nphysa{\ref@jnl{Nucl.~Phys.~A}}   
\def\physrep{\ref@jnl{Phys.~Rep.}}   
\def\physscr{\ref@jnl{Phys.~Scr}}   
\def\planss{\ref@jnl{Planet.~Space~Sci.}}   
\def\procspie{\ref@jnl{Proc.~SPIE}}   

\makeatother

\newcommand{\suzaku}{{\it Suzaku}}
\newcommand{\xmm}{{\it XMM}}

\newcommand{\nustar}{\textit{NuSTAR}}

\newcommand{\fluxcgs}{erg~s$^{-1}$~cm$^{-2}$}
\newcommand{\lumcgs}{erg~s$^{-1}$}
\newcommand{\phflux}{ph~s$^{-1}$~cm$^{-2}$}

\title[The \nustar\ view of the true Type 2 Seyfert NGC~3147]{The \nustar\ view of the true Type 2 Seyfert  NGC~3147}

\author[Stefano Bianchi, et al.]{Stefano Bianchi$^1$\thanks{E-mail: bianchi@fis.uniroma3.it (SB)}, Andrea Marinucci$^1$, Giorgio Matt$^1$, Riccardo Middei$^1$,  
\newauthor Xavier Barcons$^{2}$, Loredana Bassani$^{3}$, Francisco J. Carrera$^{2}$, Fabio La Franca$^1$,
\newauthor Francesca Panessa$^{4}$\\
$^1$Dipartimento di Matematica e Fisica, Universit\`a degli Studi Roma Tre, via della Vasca Navale 84, 00146 Roma, Italy\\
$^2$ Instituto de F\'isica de Cantabria (CSIC-Universidad de Cantabria), 39005 Santander, Spain\\
$^3$ INAF-IASF, Via P. Gobetti 101, 40129 Bologna, Italy\\
$^4$ INAF/Istituto di Astrofisica e Planetologia Spaziali, Via Fosso del Cavaliere, 00133 Roma, Italy\\
}

\begin{document}

\maketitle

\label{firstpage}

\begin{abstract}
We present the first \nustar\ observation of a `true' Type 2 Seyfert galaxy. The 3-40 keV X-ray spectrum of NGC~3147 is characterised by a simple power-law, with a standard $\Gamma\sim1.7$ and an iron emission line, with no need for any further component up to $\sim40$ keV. These spectral properties, together with significant variability on time-scales as short as weeks (as shown in a 2014 \textit{Swift} monitoring campaign), strongly support an unobscured line-of-sight for this source. An alternative scenario in terms of a Compton-thick source is strongly disfavoured, requiring an exceptional geometrical configuration, whereas a large fraction of the solid angle to the source is filled by a highly ionised gas, whose reprocessed emission would dominate the observed luminosity. Moreover, in this scenario the implied intrinsic X-ray luminosity of the source would be much larger than the value predicted by other luminosity proxies, like the [\ion{O}{iii}]$\lambda5007$ emission line extinction-corrected luminosity. Therefore, we confirm with high confidence that NGC~3147 is a true Type 2 Seyfert galaxy, intrinsically characterised by the absence of a BLR.
\end{abstract}


\section{Introduction}

NGC~3147 \citep[z=0.009346:][]{Epinat2008} is one of the three `true' Type 2 Seyfert galaxies confirmed by simultaneous optical and X-ray observations \citep{Bianchi2008c,Panessa2009b,Shi2010a,Tran2011,Bianchi2012}. These sources have a typical Type 2 optical/UV spectrum, thus lacking, by definition, broad emission lines, but this is due to the intrinsic lack of the Broad Line Region (BLR), rather than its obscuration, since the nucleus is \textit{simultaneously} seen unobscured in the X-rays. Several theoretical models predict that the BLR cannot form at very low accretion rates/luminosities \citep{Nicastro2000,Elitzur2006,Trump2011}, although a dependence on Black Hole mass somewhat relaxes these limits \citep{Elitzur2016}. All the confirmed true Type 2 Seyfert galaxies (NGC 3147 included) have indeed low accretion rates \citep{Bianchi2012}.

Previous high-throughput X-ray observations of NGC~3147 performed with \xmm-Newton and \suzaku\ put tight constraints on the column density of any obscuring material along the line-of-sight, showing the typical spectrum of an unobscured Seyfert galaxy \citep{Bianchi2008c,Matt2012}. However, these data-sets are still consistent with 
a rather extreme Compton-thick scenario in which reflection is mostly dominated by a highly ionised mirror. In this scenario, the lack of broad optical lines would be due to absorption, as in standard Unification Models \citep{Antonucci1993}, while the lack of X-ray absorption would be only apparent and due to the limited bandpass of the telescopes. The \suzaku\ observation was indeed designed to look for an excess at energies above 10 keV, a clear sign of the intrinsic continuum piercing through a Compton-thick obscurer. However, the large uncertainties in the background subtraction for the HXD/PIN spectrum prevented this observation from being conclusive in ruling out the Compton-thick nature of NGC~3147 \citep{Matt2012}. 

\nustar\ is the only instrument presently available which provides the needed high energy coverage and sensitivity in order to confirm that NGC~3147 is a true Type 2 Seyfert galaxy, and exclude the only alternative, that it may instead be a very peculiar Compton-thick source. In this paper, we therefore report on the first \nustar\ observation of NGC~3147.

\section{Observations and data reduction}

\nustar\ \citep{Harrison2013} observed NGC~3147 with its two modules (FPMA and FPMB) on 2015 December
27 for a total of $\sim79$ ks of elapsed time. The level 1 data products were processed with the \nustar\ Data Analysis Software (\textsc{nustardas}) package (v. 1.6.0). Cleaned event files (level 2 data products) were produced and calibrated using standard filtering criteria with the \textsc{nupipeline} task and the latest calibration files available in the \nustar\ calibration data-base at the time of the analysis (\textsc{caldb} 20161021). The use of more restrictive background cuts (\textsc{SAAMODE=optimized/strict}) did not yield any significant difference.
Both extraction radii for the source and background spectra were 70 arcsec, in order to maximise the signal-to-noise ratio. The resulting spectra, with an exposure time of $\sim49$ ks, were binned in order to oversample the instrumental resolution by at least a factor of 2.5 and to have a signal-to-noise ratio greater than 5 in each spectral channel. The spectra of the two modules are in good agreement with each other, the cross-normalization constant for FPMB with respect to FPMA in all the fits being within 3\%. A \textit{Swift} XRT observation of NGC~3147 was taken simultaneously to the \nustar\ observation, but, due to its limited exposure ($\sim1.5$ ks), it does not give any useful constraints, so it will not be used in the fits presented in the next section.  

In the following, errors and upper limits correspond to the 90 per cent confidence level for one interesting parameter, unless otherwise stated, apart from the plots, where always $1\sigma$ error bars are shown. The adopted cosmological parameters are $H_0=70$ km s$^{-1}$ Mpc$^{-1}$, $\Omega_\Lambda=0.73$ and $\Omega_m=0.27$ \citep[i.e. the default ones in \textsc{xspec 12.9.0}:][]{Arnaud1996}.

\section{Data Analysis}

\subsection{The \nustar\ spectrum}\label{nustaranalysis}

\begin{figure}
\begin{center}
\epsfig{file=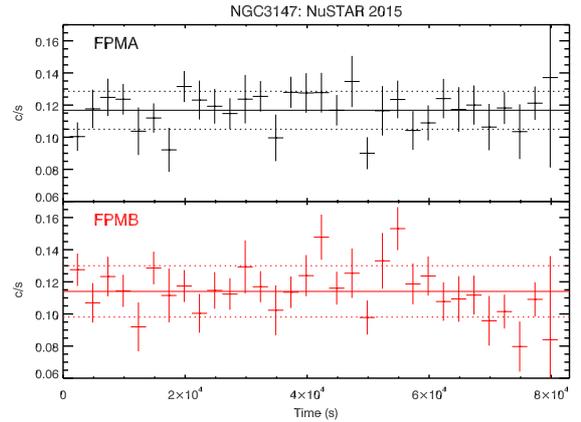,width=\columnwidth}
\end{center}
\caption{\label{3147lc}NGC~3147: \nustar\ FPMA (\textit{top}) and FPMB (\textit{bottom}) background-subtracted light-curves in the 3-78 keV energy range, for a time-bin of 2500 s. The continuous and dotted lines represent the average and one standard deviation.}
\end{figure}

As a preliminary step, we show in Fig.~\ref{3147lc} the light-curves for both modules, in the 3-78 keV energy band. We tested against the hypothesis that the source is constant via a $\chi^2$ test: in neither module the variability is significant. Therefore, in the following, we will always use the whole time-averaged spectra. Moreover, since the background becomes dominant at around 40 keV, all the fits will be performed up to that energy.

\begin{figure}
\begin{center}
\epsfig{file=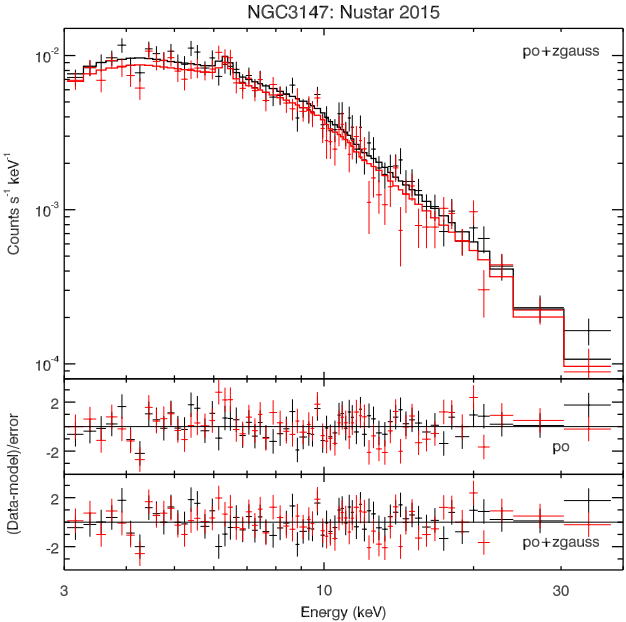,width=\columnwidth}
\end{center}
\caption{\label{3147_spectrum}\textit{Top}: \nustar\ FPMA (\textit{black}) and FPMB (\textit{red}) spectra and best-fit model. \textit{Middle}: Residuals (in $\sigma$) with respect to a model with a simple power-law. \textit{Bottom}: Residuals (in $\sigma$) with respect to a model with a power-law and a Gaussian emission line at 6.4 keV. See text for details.}
\end{figure}

The spectra are then fitted with a simple power-law (Galactic absorption does not significantly affect the data due to the limited bandpass of \nustar\ at lower energies). A photon index $\Gamma=1.70\pm0.05$ gives a good fit ($\chi^2=142/131$ d.o.f.), with few residuals left (see Fig.~\ref{3147_spectrum}). Significant residuals, in particular for FPMB, are still present around the iron line energy: the addition of an unresolved emission line at 6.4 keV improves the fit ($\chi^2=132/130$ d.o.f.: we will refer to this model as the baseline model hereinafter), yielding a flux of $3.6\pm1.9\times10^{-6}$ \phflux\ and an equivalent width EW=$120\pm60$ eV, for a $\Gamma=1.69\pm0.05$, fully consistent with the values found in past \textit{XMM-Newton} and \textit{Suzaku} observations \citep{Bianchi2008c,Matt2012}. A second emission line either at 6.7 or at 6.966 keV, corresponding to \ion{Fe}{xxv} or \ion{Fe}{xxvi} K$\alpha$ emission, as detected in the \textit{XMM-Newton} and \textit{Suzaku} data \citep[][see also next section for details]{Bianchi2008c,Matt2012}, is not required, with an upper limit to their flux of $2\times10^{-6}$ \phflux (EW$<70$ eV) and $1.3\times10^{-6}$ \phflux (EW$<50$ eV), respectively, consistent with previous detections. The 3-10 (10-78) keV flux for the baseline model is $2.27\pm0.07$ ($6.4\pm0.4$) $\times10^{-12}$ \fluxcgs, corresponding to a luminosity of $4.39\pm0.15\times10^{41}$ ($1.20\pm0.09\times10^{42}$) \lumcgs.

Although the data do not require any extra component with respect to the simple baseline model presented above, we also tried some more complex physically-motivated scenarios. We began by replacing the Gaussian emission line with a self-consistent model of reflection off a Compton-thick slab \citep[\textsc{pexmon}:][]{Nandra2007}. The fit is statistically equivalent with respect to the baseline model ($\chi^2=133/130$ d.o.f.). The intrinsic power-law index is now steeper, although still consistent within errors with the best fit ($\Gamma=1.81\pm0.08$), in order to allow for the presence of a moderate reflection component ($R=0.4^{+0.3}_{-0.2}$), the inclination angle being fixed at $30\degr$. If a \textsc{cutoffpl} is used instead of a power-law, no high-energy cut-off can be measured, the lower limit being 50 keV.

The main objective of this \nustar\ observation was to look for a hard X-ray excess, in order to test if NGC~3147 is a Compton-thick source instead of a true Type 2 Seyfert galaxy. In this scenario the observed power-law component would be reflection (Compton scattering) from a highly ionised gas, while the \textsc{pexmon} component is reflection from a cold Compton-thick gas. The smoking gun for this alternative scenario would have been the detection of the primary emission of the source piercing above 10 keV through a Compton-thick absorber. Such a component is not present in the \nustar\ data. A basic requirement is that the (absorbed) intrinsic luminosity is at least as large as the observed one. In this case, adding to the previous model a power-law with the same photon index and luminosity as the observed one, but absorbed by a large column density (both photoelectric and Compton opacities are taken into account), a $\mathrm{N_H}>3\times10^{24}$ cm$^{-2}$ is needed in order for this component not to be observed in our data. Assuming a more reasonable ratio of 15 between the intrinsic and the observed luminosity, the required column density becomes $\mathrm{N_H}>6\times10^{24}$ cm$^{-2}$. We will discuss this scenario extensively in Sect.~\ref{discussion}.

\subsection{Previous X-ray observations of NGC~3147}

\begin{figure}
\begin{center}
\epsfig{file=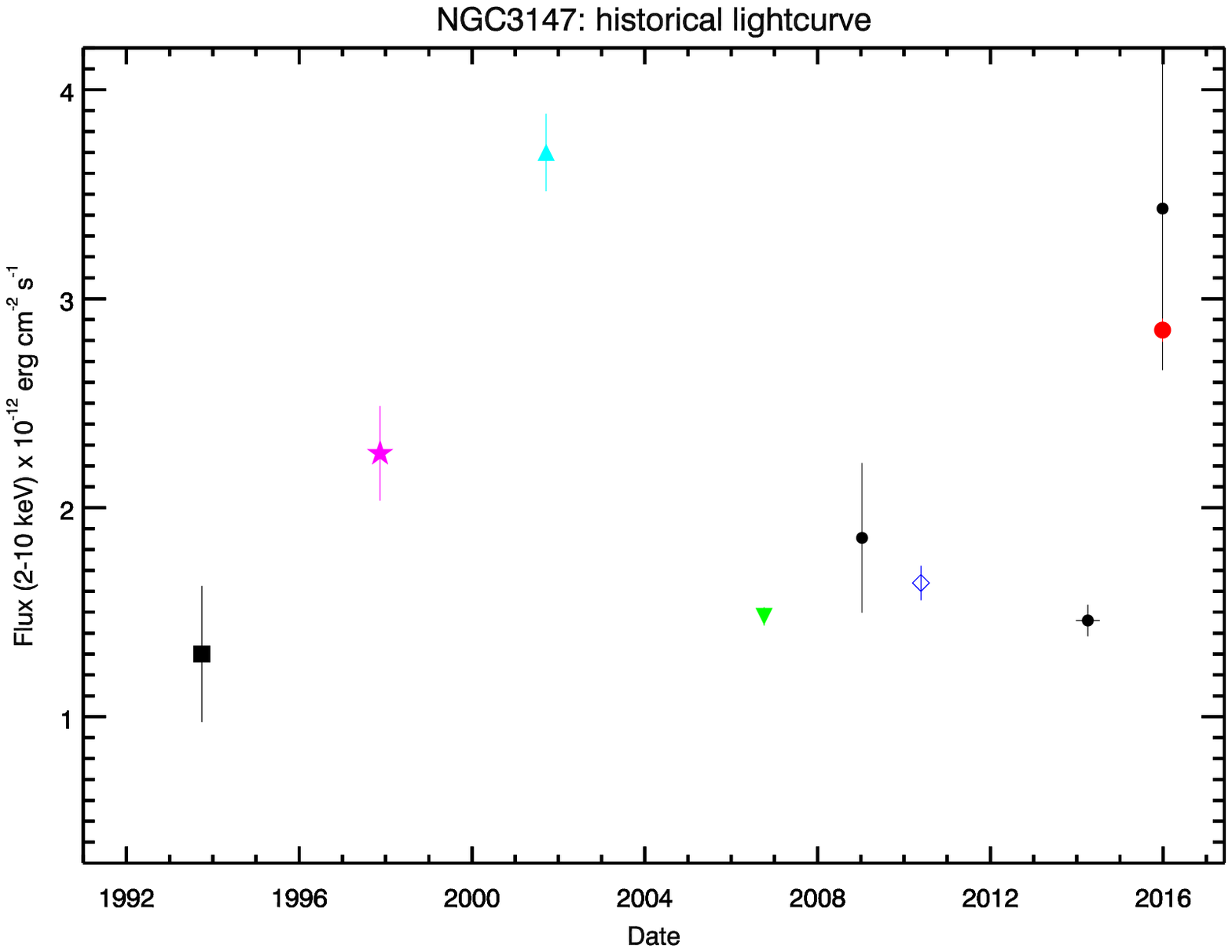,width=\columnwidth}
\end{center}
\caption{\label{fluxhistory}2-10 keV flux of all past X-ray observations of NGC~3147: long-term variability is evident. Black filled square: \textit{ASCA} \citep{Ptak1996}; magenta filled star: \textit{BeppoSAX} \citep{Dadina2007}; cyan filled triangle: \textit{Chandra} \citep{Terashima2003a}; green upside down triangle: \textit{XMM-Newton} \citep{Bianchi2008c}; blue open diamond: \textit{Suzaku} \citep{Matt2012}; black filled circle: \textit{Swift} (this paper); red filled circle \nustar\ (this paper). The 2014 \textit{Swift} flux is the weighted mean of the observations of the campaign plotted in Fig.~\ref{swiftlc}. Error bars taken from the same references, apart from \textit{BeppoSAX}, \textit{Chandra} and \textit{Suzaku}, for which we assumed 10\%, 5\% and 5\% errors.}
\end{figure}

NGC~3147 has an extensive record of X-ray observations. In Fig.~\ref{fluxhistory} we plot all the observed 2-10 keV fluxes of the sources, from the 1993 \textit{ASCA} observation to the 2015 \nustar\ observation presented in this work. A long-term variability up to a factor of $\sim2$ is evident. It should be noted that the higher flux is the one measured with the telescope with the better spatial resolution (\textit{Chandra}), thus excluding that the variability is driven by unresolved off-nuclear sources in the host galaxy. Variability in the soft X-ray energy band was also reported between the \textit{XMM-Newton} and \textit{Suzaku} observations by \citet{Matt2012}.

\begin{figure}
\begin{center}
\epsfig{file=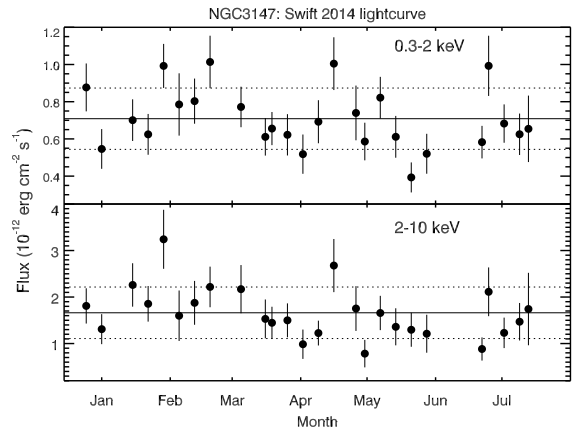,width=\columnwidth}
\end{center}
\caption{\label{swiftlc}The \textit{Swift} 2014 monitoring campaign of NGC~3147. Soft (\textit{top}) and hard (\textit{bottom}) fluxes are plotted for all the observations of the campaign, together with their average and standard deviation. The light-curves are significantly variable in both energy bands. See text for details.}
\end{figure}

A weekly cadenced, six-month monitoring campaign of NGC~3147 was carried out in 2014 by \citet{Andreoni2016} with \textit{Swift} XRT. We re-extracted and fitted all the spectra of these observations, together with the two other available 2009 and 2015 observations, with a simple power-law absorbed by the Galactic column density \citep[$3.64\times10^{20}$ cm$^{-2}$][]{Dickey1990}. The resulting 0.3-2 keV and 2-10 keV fluxes are plotted in Fig.~\ref{swiftlc}. We tested against the hypothesis that the source is constant via a $\chi^2$ test: in both energy bands the light-curve is significantly variable (99\% and 97\% confidence level for the soft and the hard band, respectively). NGC~3147 is therefore variable also on shorter time-scales, of the order of weeks.

Before the \textit{Suzaku} observation, NGC~3147 was not detected above 10 keV, with only upper limits to its flux: $F_{20-100~\mathrm{keV}}<1.3\times10^{-11}$ \fluxcgs\ \citep[\textit{BeppoSAX}, 3$\sigma$:][]{Dadina2007}, $F_{20-40~\mathrm{keV}}<4.5\times10^{-12}$ \fluxcgs\ and $F_{40-100~\mathrm{keV}}<8.7\times10^{-12}$ \fluxcgs\ \citep[\textit{INTEGRAL}, 3$\sigma$:][]{Bird2016}, and $F_{15-150~\mathrm{keV}}<4\times10^{-12}$ \fluxcgs\ \citep[\textit{Swift}, 3$\sigma$:][]{Matt2012}. The \textit{Suzaku} spectrum yielded nominally a flux $F_{15-100~\mathrm{keV}}\sim1.2\times10^{-11}$ \fluxcgs\, but the uncertainties on the background level made it consistent with the \textit{Swift} BAT upper limit \citep{Matt2012}. Indeed, our \nustar\ data have a 20-40 keV flux of $\sim2.1\times10^{-12}$ \fluxcgs\, well below the \textit{INTEGRAL} upper limit, supporting the interpretation that the \textit{Suzaku} HXD/PIN flux was heavily contaminated by the background. On the other hand, the extrapolation of our \nustar\ best fit to 150 keV gives a flux $F_{15-150~\mathrm{keV}}\sim8\times10^{-12}$ \fluxcgs\, which is larger than the \textit{Swift} BAT upper limit. The latter refers to the 66-month Palermo BAT Catalogue \citep{Segreto2010}, so it is an average flux in the years 2005-2010. As it can be seen from Fig.~\ref{fluxhistory}, the average 2-10 keV flux of NGC~3147 in that period was roughly half of that measured by \nustar: a variability of the same order at higher energies is enough to explain the discrepancy between \nustar\ and \textit{Swift} BAT, but also the possible presence of a cut-off at energies above $\sim50$ keV can make a significant contribution.

A strong neutral iron emission line was detected by \textit{ASCA}, with an EW$=500\pm300$ eV \citep{Ptak1996}. This feature was later confirmed by \textit{XMM-Newton} and \textit{Suzaku}, with EWs of  $130\pm80$ and $200\pm40$ eV, respectively \citep{Bianchi2008c,Matt2012}.  Within the (large) errors, all these values are consistent with the EW$=120\pm60$ eV measured in the \nustar\ spectrum, although larger EWs tend to correspond to lower 2-10 keV fluxes (see Fig.~\ref{fluxhistory}). Indeed, the observed fluxes of the line at 6.4 keV are also consistent with each other, so there is no evidence against a constant emission line over long time-scales. 

Another emission line at higher energies was also detected by \textit{ASCA}, \textit{XMM-Newton} and \textit{Suzaku}, with flux and EW always comparable to that of the neutral iron line. The data-set with the best signal-to-noise ratio, i.e. \textit{Suzaku}, preferred an identification as \ion{Fe}{xxvi} K$\alpha$ emission. Such a feature is not required in the \nustar\ data, but the upper limit to its flux is consistent with previous detections, indicating that also this emission line may be constant.

\section{Discussion}\label{discussion}

The \nustar\ spectrum of NGC~3147 shows no significant deviations from an unabsorbed power-law up to 40 keV, with a photon spectral index $\Gamma\simeq1.7$, in very good agreement with the average value found in local Seyfert galaxies \citep[e.g.][]{Bianchi2008}. Long- and short-term (down to weeks) variability, both in the soft and in the hard X-ray energy bands, is also in agreement with an identification of this object as a standard unabsorbed Seyfert galaxy. Therefore, given the simultaneous lack of X-ray absorption and of broad optical lines \citep{Bianchi2008c}, it seems inescapable to conclude that NGC~3147 is indeed a true Type 2 Seyfert galaxy, whose lack of broad permitted line components in their observed optical spectrum is not due to reddening along the line of sight, but to the intrinsic absence of the BLR \citep[e.g.][]{Panessa2002,Bianchi2012}. As extensively discussed in \citet{Bianchi2008c} and \citet{Bianchi2012}, it is possible that the very low accretion rate of this source prevents the formation of the BLR \citep[e.g.][]{Nicastro2000,Trump2011}.

However, as discussed by \citet{Matt2012} and confirmed in Sect.~\ref{nustaranalysis}, the X-ray spectrum is still compatible with a Compton-thick source, whose line-of-sight is completely blocked up to very high energies by neutral material, but whose observed spectrum is dominated by reflection from a highly ionised `mirror'. The observed 2-10 keV luminosities of the warm (power-law) and cold (\textsc{pexmon}) reflection components, as in the fit presented in Sect.~\ref{nustaranalysis}, are $L_w=5.4\times10^{41}$ and $L_c=2\times10^{40}$ \lumcgs. In this scenario, $L_w$ depends on the Compton optical depth $\tau\equiv\mathrm{N_H^w}/(1.5\times10^{24} \mathrm{cm}^{-2})$ and the covering factor $f\equiv\Omega/4\pi$ of the warm reflector. In the Compton-thin regime, the relation between $L_w$ and the intrinsic luminosity of the source $L_i$ is approximately linear:

\begin{equation}
L_w \simeq L_i \tau f.
\end{equation}

On the other hand, the reflected luminosity $L_c$ from a neutral Compton-thick gas will be proportional to the intrinsic luminosity and the covering factor of the gas. Taking $R\equiv\Omega/2\pi$ as customary in reflection models (typically computed for slab or disk geometries):

\begin{equation}
L_c=L_i k R/2.
\end{equation}

In the case of \textsc{pexmon}, assuming $\Gamma=1.8$ and an inclination angle $i=30\degr$, $k=0.09$.
Finally, we can assume for simplicity that all the available solid angle is filled by matter, so that:

\begin{equation}
f+R/2=1.
\end{equation}

Therefore, the intrinsic luminosity of NGC~3147 in this scenario can be derived from the above relations, yielding:

\begin{equation}
L_i=\frac{L_w}{\tau} + \frac{L_c}{k} = \left(8.1~\mathrm{N_H^w}_{23}^{-1} + 0.2\right) \times 10^{42} ~\mathrm{erg~s}^{-1}.
\end{equation}

\begin{figure}
\begin{center}
\epsfig{file=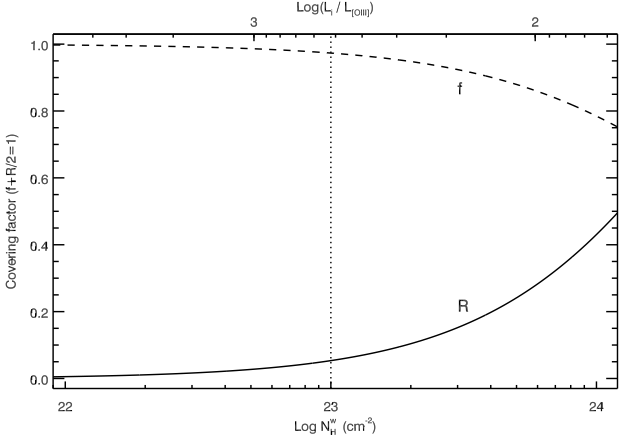,width=\columnwidth}
\end{center}
\caption{\label{scenariothick}Dependence of the covering factors of the warm mirror ($f$) and the neutral reflector ($R$) as a function of the column density of the warm mirror, in the Compton-thick scenario. The needed intrinsic luminosity of the source is also reported in the upper axis, re-scaled with respect to the observed [\ion{O}{iii}]$\lambda5007$ emission line extinction-corrected luminosity: note that the the average value of this ratio for Seyfert galaxies ($\sim1.09$ in log) is outside the plot. The vertical dotted line represents the example described in the text.}
\end{figure}

The allowed solutions for $R$, $f$, $\mathrm{N_H^w}$ and $L_i$ are shown in Fig.~\ref{scenariothick}. In particular, large covering factors for the warm reflector are always needed, with $R$ becoming vanishingly small when the column density of the warm mirror falls below $\sim10^{23}$ cm$^{-2}$. For example, for $\mathrm{N_H^w}=10^{23}$ cm$^{-2}$, we need an intrinsic luminosity $L_i=8.3\times10^{42}$ \lumcgs, $f\simeq0.97$ and $R\simeq0.05$. With these values, $L_i/(L_w+L_c)\simeq15$, so that a neutral absorbing column density of at least $6\times10^{24}$ cm$^{-2}$ should intercept our line of sight in order to completely suppress it in our \nustar\ data (see Sect.~\ref{nustaranalysis}). Given the very low value of $R$ derived in this scenario, this neutral Compton-thick gas should be only in the line of sight, while almost all the available solid angle should be filled by the warm mirror.

This geometrical configuration would be rather exceptional, and never observed so far in a Compton-thick AGN. NGC~1068 is one of the Compton-thick sources with the most prominent contribution from a highly ionised reflector. However, even in this source the ratio between the warm and the cold reflection luminosities is only $\sim0.7$ \citep[][]{Matt2004}, much lower than the value ($\sim27$) that we would have in NGC~3147. Moreover, the variability observed on a time-scale of weeks (see Fig.~\ref{swiftlc}) implies that at least the dominant component (the warm reflector) should be very compact. Even if being mostly along the line-of-sight, the absorber (and cold reflector) must then be also very compact, in order to leave the warm reflector observable. However, in order to hide also the BLR, it should be at least outside the sublimation radius, which in this source is of the order of $r_s\sim 1.3 L_{UV,46}^{1/2}\simeq 0.03$ pc \citep[assuming $L_{UV}\sim L_{bol}$:][]{Barvainis1987,Bianchi2008c}. 

Finally, the intrinsic luminosity of the Compton-thick scenario depicted above should be compared to other luminosity proxies, like the [\ion{O}{iii}]$\lambda5007$ emission line extinction-corrected luminosity, which is $1.6\times10^{40}$ \lumcgs\ \citep{Bianchi2008c}. For a column density of the warm mirror of $10^{23}$ cm$^{-2}$, the ratio $\log\dfrac{L_i}{L^c_\ion{O}{iii}}\simeq2.7$ is much larger (a factor of $\sim40$) than the average value for Seyfert galaxies \citep[$1.09\pm0.63$:][]{Lamastra2009}. A column density of the warm mirror larger than $10^{24}$ cm$^{-2}$ is required to have an intrinsic luminosity low enough to allow for this luminosity ratio to be within the dispersion of the observed distribution (see Fig.~\ref{scenariothick}). On the other hand, we note that the \textit{observed} luminosity of NGC~3147, which corresponds to its intrinsic luminosity in the preferred unabsorbed scenario, gives immediately a ratio $\log\dfrac{L_o}{L^c_\ion{O}{iii}}\simeq1.5$, in agreement with the distribution of local unabsorbed or Compton-thin Seyfert galaxies \citep[Compton-thick Seyfert galaxies have $\sim10-100$ times smaller values:][]{Bassani1999}.

\section{Conclusions}

We presented the analysis of the first \nustar\ observation of a true Type 2 Seyfert galaxy. The spectrum of NGC~3147 can be simply modelled by a power-law with a standard $\Gamma\sim1.7$ and an iron emission line, with no need for any further component even at energies larger than 10 keV. These spectral properties, together with significant variability on time-scales as short as weeks, strongly support a line-of-sight free of absorption for this source.

An alternative scenario in terms of a Compton-thick source cannot be strictly ruled out from a purely spectroscopic point of view. However, this scenario requires an exceptional geometrical configuration, where a large fraction of the solid angle to the source would be filled by a highly ionised gas, whose reprocessed emission would dominate the observed luminosity. A neutral Compton-thick gas would completely absorb the intrinsic emission along the line-of-sight, while its contribution to the overall reprocessed emission would be almost negligible. Moreover, the intrinsic X-ray luminosity of the source required by this scenario would be much larger than expected, with respect to other luminosity proxies, like the [\ion{O}{iii}]$\lambda5007$ emission line extinction-corrected luminosity.

Therefore, the \nustar\ data adds further evidence in favour of an X-ray spectrum completely unaffected by absorption, confirming NGC~3147 as one of the best cases of true Type 2  Seyfert galaxies, intrinsically characterised by the absence of a BLR.

\section*{Acknowledgements}

We would like to thank Kristin Kruse Madsen for her help in interpreting the discrepancy at the iron line energy between the two \nustar\ modules.
SB acknowledges financial support from the Italian Space Agency under grant ASI-INAF I/037/12/0. AM and GM acknowledge financial support from the Italian Space Agency under grant ASI/INAF I/037/12/0-011/13. AM acknowledges financial support from the European Union Seventh  Framework Programme (FP7/2007-2013) under grant agreement no. 312789. FP acknowledges the ASI/INAF agreement number 2013-023-R1. XB and FJC acknowledge financial support through grant AYA2015-64346-C2-1-P (MINECO/FEDER).

\bibliographystyle{mnras}
\bibliography{NGC3147}

\label{lastpage}

\end{document}